\documentclass[twocolumn,showpacs,amsmath,amssymb]{revtex4}
\usepackage{graphicx}
\usepackage{color}
\usepackage{amsmath}
\usepackage{amssymb}
\usepackage{booktabs}
\usepackage{float}

\begin{document}
\title{Experimental Realization of the Green-Kubo Relation in Colloidal  Suspensions Enabled by Image-based Stress Measurements}
\author{Neil Y.C. Lin}
\affiliation{Department of Physics, Cornell University, Ithaca, New York 14853}
\author{Matthew Bierbaum}
\affiliation{Department of Physics, Cornell University, Ithaca, New York 14853}
\author{Itai Cohen}
\affiliation{Department of Physics, Cornell University, Ithaca, New York 14853}

\begin{abstract}
By combining confocal microscopy and Stress Assessment from Local Structural Anisotropy (SALSA), we directly measure stresses in 3D quiescent colloidal liquids. Our non-invasive and non-perturbative method allows us to measure forces $\lesssim$ 50 fN with a small and tunable probing volume, enabling us to resolve the stress fluctuations arising from particle thermal motions. We use the Green-Kubo relation to relate these measured stress fluctuations to the bulk Brownian viscosity at different volume fractions and comparing against simulations and conventional rheometry measurements. We demonstrate that the Green-Kubo analysis gives excellent agreement with these prior results. This agreement provides a strong demonstration of the applicability of the Green-Kubo relation in nearly hard-sphere suspensions and opens the door to investigations of local flow properties in many poorly understood far-from-equilibrium systems, including suspensions that are glassy, strongly-sheared, or highly-confined.
\end{abstract}


\maketitle

All quiescent thermal systems may seem static macroscopically, but microscopically they fluctuate strongly. By observing the system's response to these thermal fluctuations, a material's linear transport coefficients can be predicted using the Green-Kubo relation~\cite{kubo1957statistical, green1954markoff, morriss2007statistical, hansen2013theory}. This foundational relation -- a central achievement of nonequilibrium statistical mechanics -- has enabled numerous diverse theoretical calculations ranging from electrical and magnetic susceptibilities in quantum systems~\cite{baranger1989electrical, alhassid2000statistical} to thermal conductivities in nanotubes~\cite{berber2000unusually, savin2009thermal, yao2005thermal, zhang2004chirality}. In particular, it has been widely used to theoretically determine the viscosities in bulk~\cite{hoover1980lennard}, confined~\cite{neek2016commensurability, huang2014green}, supercooled~\cite{kushima2009computing, sosso2012breakdown}, and quantum~\cite{reichman2001self} liquids, where external load is problematic or heterogeneities play a crucial role. Unfortunately, these applications have remained strictly theoretical due to the difficulties in experimentally observing fluctuations in atomic systems, which are too rapid ($\sim$ ps) and weak ($\sim$ $\mu$N) to mechanically resolve in experiments.

\begin{figure} [htp]
\includegraphics[width=0.5 \textwidth]{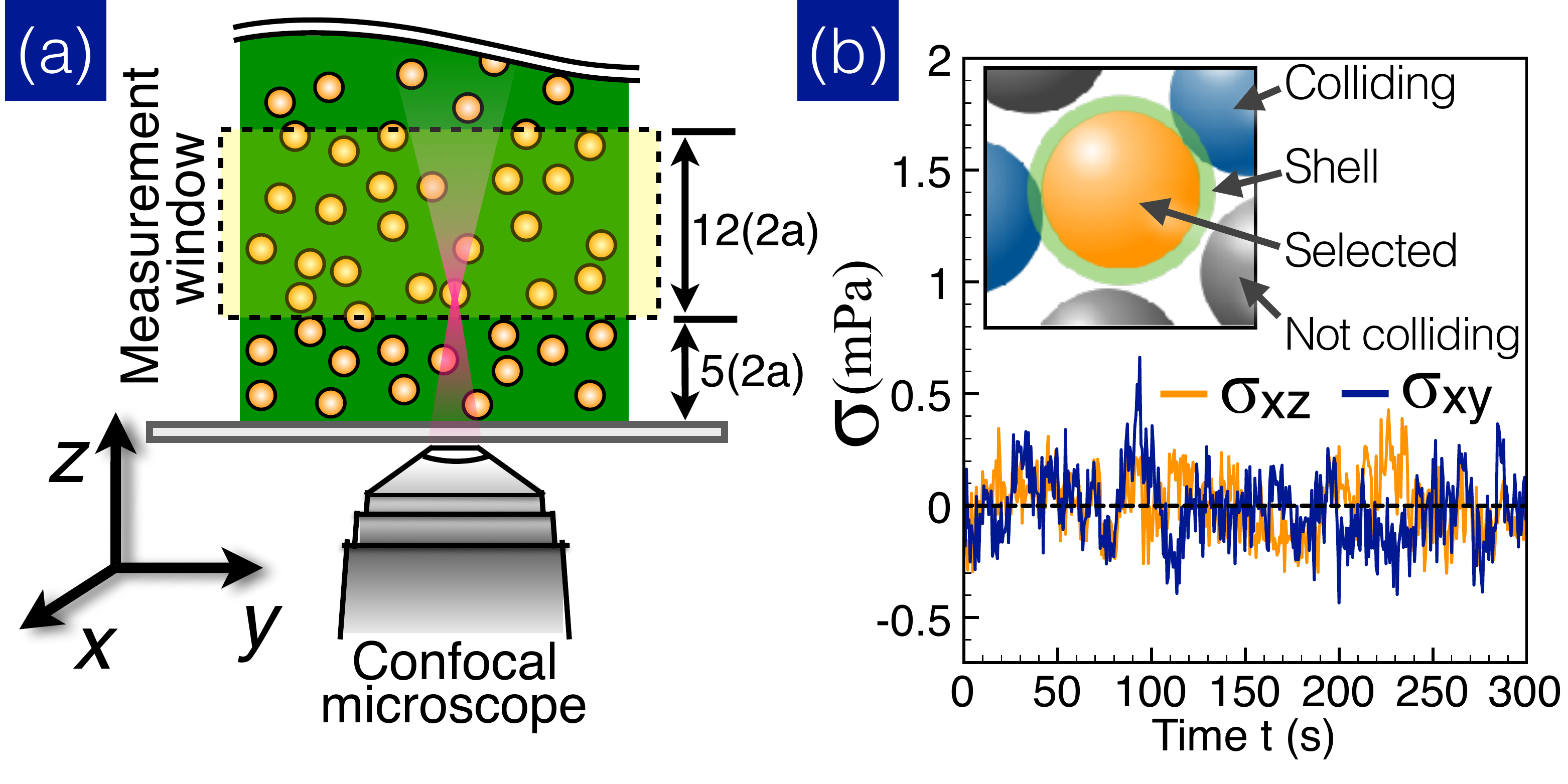}
\caption{(a) Schematic of the experiment setup and axis. The suspension sample is hermetically sealed in a sample cell, and placed on a high-speed confocal microscope to image its microstructure. The measurement window (dashed box) is $\sim5$ $\mu$m above the coverslip avoiding boundary effects. (b) The featured particle positions are used to calculate the stress using SALSA. The selected particle's (orange) local structural anisotropy is calculated based on the configuration of its colliding neighbors (blue) that lie within a thin shell $\Delta \approx 106$ nm (green). This process is done for each snapshot giving the instantaneous Brownian stresses $\sigma_{xz}$ (orange line) and $\sigma_{xy}$ (blue line) fluctuating within $\sim \pm$ 0.5 mPa.}
\label{fig_1}
\end{figure}

Here, by using high-speed confocal microscopy in conjunction with \textit{Stress Assessment from Local Structural Anisotropy} (SALSA)~\cite{lin2016measuring}, we directly measure the stress fluctuations in nearly hard-sphere colloidal liquids. Colloidal suspensions are comprised of particles that are small enough to demonstrate Brownian motions, while large enough to be optically imaged, providing length- and time-scales that are associated with system relaxation~\cite{foss2000structure}. To measure a suspension's stress fluctuations, we use a confocal microscope to image the 3D microstructure of the sample, then use SALSA to determine its Brownian stress arising from interparticle thermal collisions. Since SALSA is image-based, non-invasive, non-perturbative, and able to measure the suspension stress with a tunable probing volume, it can resolve the weak stress fluctuations that are usually averaged out in conventional bulk measurements due to the requisite large probing volume.

\begin{figure*} [htp]
\includegraphics[width=0.9 \textwidth]{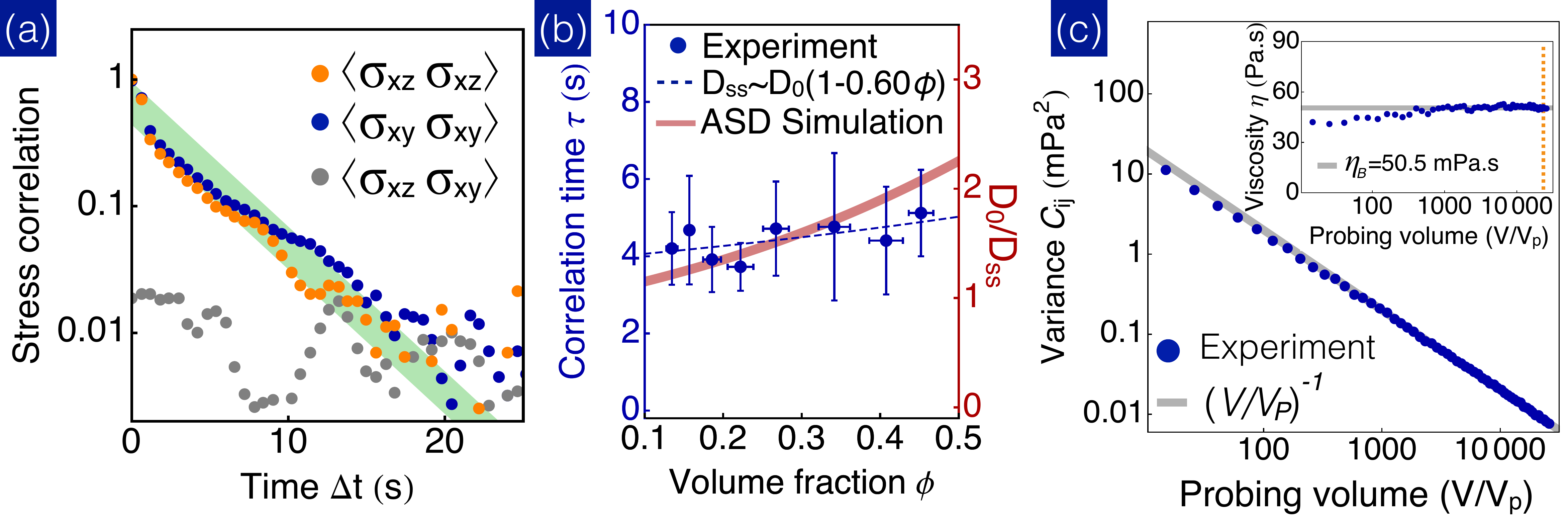}
\caption{(a) Time-time autocorrelation functions of $\sigma_{xz}$ (orange) and $\sigma_{xy}$ (blue) are calculated from the time series of stress. The green line shows an exponential fit to the data to extract the time scale of the stress autocorrelation functions. The cross-correlation $\langle \sigma_{xz}(t + \Delta t) \sigma_{xy}(t)\rangle$ (gray) is consistent with zero showing low coupling between components. For clarity, the autocorrelation is normalized by its corresponding fluctuation's variance, and the cross-correlation is normalized by the mean variance of all autocorrelations. (b) The correlation time $\tau$ of the stress fluctuation varies weakly with the volume fraction $\phi$, a trend that is consistent with the variation of the inverse self-diffusivity $D_0/D_{ss}(\phi)$ found in Accelerated Stokesian Dynamics simulations (ASD)~\cite{Banchio2003}. Here, $D_0$ is the diffusivity in ultra-dilute limit and $a^2 / D_{ss}$ roughly determines the relaxation time-scale of the suspension. (c) Mean variance, $C_{ij}$, of all shear stress components is plotted versus the normalized probing volume $V/V_{p}$, where $V_p$ is particle volume. The gray line denotes an inverse proportionality between $C_{ij}$. The inset shows that the measured viscosity $\eta\approx 50.5$ mPa.s is roughly constant when $V/V_{p}\ge200$ and starts to decay slightly at smaller probing volumes. We set the measurement window $V = 61\mu$m$\times15\mu$m$\times12\mu$m, $V/V{p}\sim 22,280$ (orange line) throughout all measurements.}
\label{fig_2}
\end{figure*}

The suspension samples are comprised of silica spheres with a radius $a=$ 490 nm in a water-glycerine mixture that has a matched refractive index and viscosity $\eta_0=60$ mPa$\cdot$s. We add 1.25 mg/ml of fluorescein sodium salt to the solvent to shorten screening length ($\le$10~nm) and obtain nearly hard sphere interactions. The added fluorescein also makes the solvent fluorescent, so the solvent appears bright and the particles appear  dark. We then image the particle configuration using a high-speed confocal microscope with a hyper-fine scanner that maximizes the stability in the vertical ($z$-axis) scanning position (schematic in Fig.~\ref{fig_1}(a)). To ensure that the suspension structure remains homogenous throughout the experiment, we image the sample within a minute after the sample cell is made. We capture 216 frames per second and acquire stacks of 100 images within $0.5~\rm{s} \sim 0.02\tau_B$, where $\tau_B=6 \pi a^3 \eta_0/k_B T$ is the self-diffusion time of the sample.

By implementing the previously developed SALSA method~\cite{lin2016measuring}, we determine the stress in our 3D suspensions. SALSA uses the featured particle positions to calculate the \textit{local structural anisotropy} or fabric tensor $\psi_{ij}^\alpha(\Delta) = \sum_{\beta \in nn} \hat{r}_i^{\alpha \beta} \hat{r}^{\alpha \beta}_{j}$ of particle $\alpha$, where $nn$ is the set of colliding neighbors that lie within a distance $2a + \Delta$ from particle $\alpha$ ($\Delta = 106$ nm in the current work), $i$,$j$ are spatial indices, and $\hat{r}_{ij}$ is the unit vector between particles (see Fig.~\ref{fig_1}(b) and SI). Scaling the ensemble-averaged $\psi_{ij}^\alpha(\Delta)$ by $\Delta$ enables us to estimate the probability of thermal collisions between particles. Consequently, the instantaneous Brownian stress of the sample can be approximated: $\sigma_{ij} (V, \Delta) = \frac{k_B T}{V} \frac{a}{\Delta} \sum_{\alpha \in V} \psi_{ij}^\alpha(\Delta) + nk_B T \delta_{ij}$, where $V$ is the averaging window volume, $k_B T$ is thermal energy, $n$ is number density, and $\delta_{ij}$ is Kronecker delta function. Here, $nk_B T$ is simply the ideal gas term.

The typical volume of our probed region $V = 61\mu$m$\times15\mu$m$\times12\mu$m $\sim 10~\rm{pL}$ contains approximately 6,000 particles at a volume fraction $\phi\sim 0.27$. This small volume ensures that the stress fluctuations are not suppressed by the volume averaging, $\propto 1/V$, while preserving bulk behavior. We plot the instantaneous stress $\sigma_{xz}$ and $\sigma_{xy}$ in Fig.~\ref{fig_1}(b), where $\hat z$ is the gravitational axis and $\hat x$ and $\hat y$ are horizontal. In contrast to a flat line at zero level anticipated in a macroscopic measurement, we find that both $\sigma_{xz}$ and $\sigma_{xy}$ fluctuate up to $\pm$ 0.5 mPa. We note that the force fluctuations corresponding to these stresses are less than 50 fN, difficult to resolve using mechanical methods.

We calculate the time-time autocorrelation function $\langle\sigma_{ij}(t+\Delta t)\sigma_{ij}(t)\rangle$ for the stress components $\sigma_{xz}$ and $\sigma_{xy}$, and show the correlation decay in a log-linear plot, see Fig.~\ref{fig_2} (a). Despite the slight sedimentation due to the density mismatch between the particle and solvent, both autocorrelation functions decay in the same fashion indicating an isotropic viscosity of the sample (see SI). We further examine the cross-correlation $\langle \sigma_{xz}(t + \Delta t) \sigma_{xy}(t)\rangle$ and find it negligibly small, which is consistent with the system symmetry. While the exact function form of the autocorrelation decay cannot be determined from the current data due to the limited measurement time span, we use an exponential decay ($\sim e^{-\Delta t/\tau}$) to quantify the correlation time. In doing this, we find that the correlation time $\tau$ varies weakly with the suspension volume fraction $\phi$ (see Fig.~2 (b)). We compare our observed trend with previous simulations of short-time diffusivity $D_{ss}$ where $a^2 / D_{ss}$ roughly sets the relaxation time-scale of the system~[19,20]. In simulations, $D_{ss}$ decays approximately as $D_{ss}\sim D_0(1- b \phi)$ (red line, Fig.~2 (b)) with $b$ on the order of 1.5 at intermediate volume fractions. Here, we find a weaker trend $b\sim0.60\pm0.23$ (blue dashed line) indicating either our measurements are not sufficiently precise to determine $b$ accurately or that the functional form changes at volume fractions approaching close-packing.

With the measured stress fluctuations, we can directly calculate the shear viscosity of our sample via the Green-Kubo formula $\eta_B =\langle \frac{V}{k_B T} \int \langle \sigma_{ij}(t+\Delta t) \sigma_{ij}(t)\rangle d\Delta t \rangle_{i\neq j}$, where $\eta_B$ is the Brownian contribution to the total shear viscosity $\eta_{tot}$. 
Since our suspension systems are nearly hard-sphere, we anticipate that the stresses are weakly correlated in space, and thus the sample viscosity is roughly independent of probe window size. To verify this, we change our probing (averaging) volume $V$, and investigate how the stress fluctuations vary. In Fig.~\ref{fig_2}(c), we plot the mean variance of shear stress $C_{ij} = \langle \sigma_{ij}(t) \sigma_{ij}(t) \rangle_{t, i \neq j}$ as a function of $V/V_p$ where $V_p$ is the particle volume $(4/3) \pi a^3$. We find that $C_{ij}$ is inversely proportional to $V/V_p$ when $V/V_p \ge 200$ corresponding to a cubic volume that is approximately six particles across. This inverse proportionality and constant viscosity shown in the inset of Fig.~\ref{fig_2}(c) are consistent with the Green-Kubo formula. When $V/V_p \le 200$, we find that the viscosity slightly deviates from its bulk value. The viscosity reduction is around $20\%$ of the mean for the smallest probing volume explored -- a three-particle wide cube. While this reduction is reminiscent of the system size-dependent viscosity associated with long-ranged stress correlations in atomic simulations~\cite{levashov2011viscosity, yeh2004system, erpenbeck1995einstein, meier2004transport, heyes2007system, chattoraj2013elastic, maloney2006amorphous}, in our nearly hard-sphere liquid system we do not anticipate such long-ranged correlations that lead to nonlocal viscosities. Instead, at small volumes, the stress fluctuations are strongly influenced by changes in particle number as particles pass into and out of the constrained field of view.


\begin{figure} [htp]
\includegraphics[width=0.5 \textwidth]{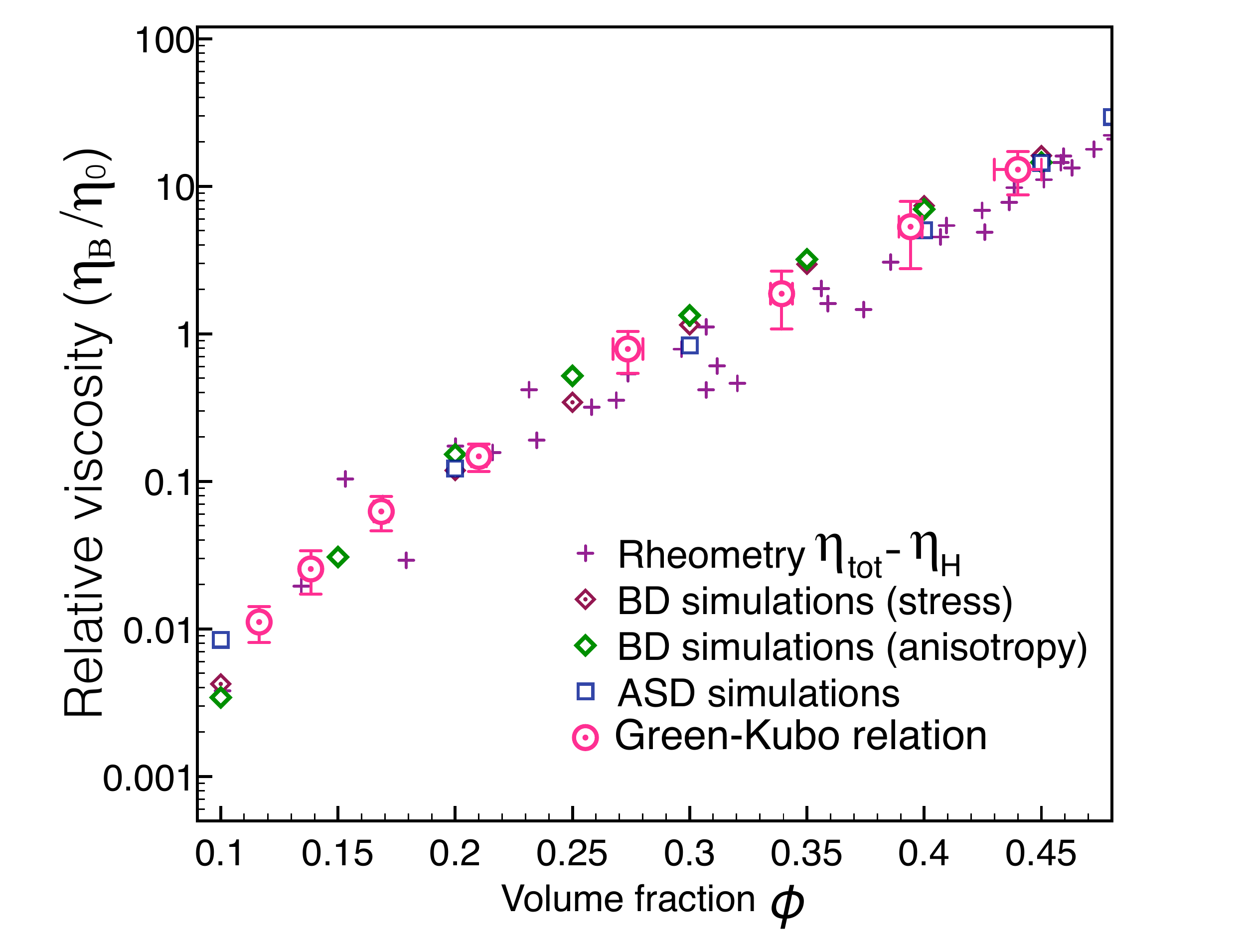}
\caption{Relative Brownian viscosity $\eta_B/\eta_0$ calculated from Grenn-Kubo relation (red circles) is plotted versus volume fraction $\phi$, where $\eta_0$ is the solvent viscosity. The error bars denote the standard errors over 14 runs of measurements. The experimental results are quantitatively consistent with accelerated Stokesian dynamics simulations, ASD (blue squares)~\cite{Banchio2003}. Furthermore, we find that the measured Brownian viscosity of our suspension is also consistent with our Brownian dynamics simulation results determined with direct stress calculation $\vec F \vec x$ (purple diamonds) and the SALSA method (green diamonds). Finally, we find our results are in excellent agreement with rheometry measurements (purple crosses)~\cite{Cheng2002}. In particular, we subtract the hydrodynamic contribution $\eta_H$ from the total viscosity $\eta_{tot}$ measured using bulk rheometry to determine the Brownian component, where $\eta_H$ is obtained from analytical calculations reported in previous work~\cite{Cheng2002, mewis2012colloidal, sierou2001accelerated, lin2014biaxial}.}
\label{fig_3}
\end{figure}

To compare our results with macroscopic flow measurements and simulations, we use the measured stress autocorrelation in conjunction with the Green-Kubo relation to determine the Brownian viscosity $\eta_B$ of suspensions at eight different volume fractions $0.12 \le \phi \le 0.45$ (see Fig.~\ref{fig_3}). The resulting viscosities (red circles) show excellent agreement with previous hydrodynamic Stokesian simulations (blue squares) ~\cite{Banchio2003}. To further confirm the accuracy of our SALSA stress measurement, we also use Brownian Dynamics simulations to generate sets of particle configurations matching the experimental parameters (\textit{e.g.} particle size, solvent viscosity, and temperature), and compare the stresses calculated from actual virials $F_{ij} X_{ij}$ (purple diamonds) with those calculated on the same data set with SALSA (green diamonds)~\cite{guazzelli2011physical}. Both results again show a quantitative agreement with the experimental measurements. Finally, the measured Brownian viscosities are compared with conventional mechanical measurements by subtracting the hydrodynamic contribution $\eta_H$ from the total viscosity $\eta_{tot}$ determined using rheometry (purple crosses)~\cite{Cheng2002, cheng2011imaging}. The rheology data points (colloidal PMMA and silica systems) are obtained from previous experiments~\cite{Cheng2002} and the hydrodynamic contribution is calculated from previous analytical approximation for the high frequency viscosity~\cite{Cheng2002, sierou2001accelerated, mewis2012colloidal, lin2014biaxial}. We find good agreement between the viscosities determined by our stress fluctuation measurements and conventional rheometry at all volume fractions explored. Collectively, the agreements between our results, simulations, and bulk measurements provide a clear demonstration of the Green-Kubo formula in hard-sphere systems.

\begin{figure} [htp]
\includegraphics[width=0.5 \textwidth]{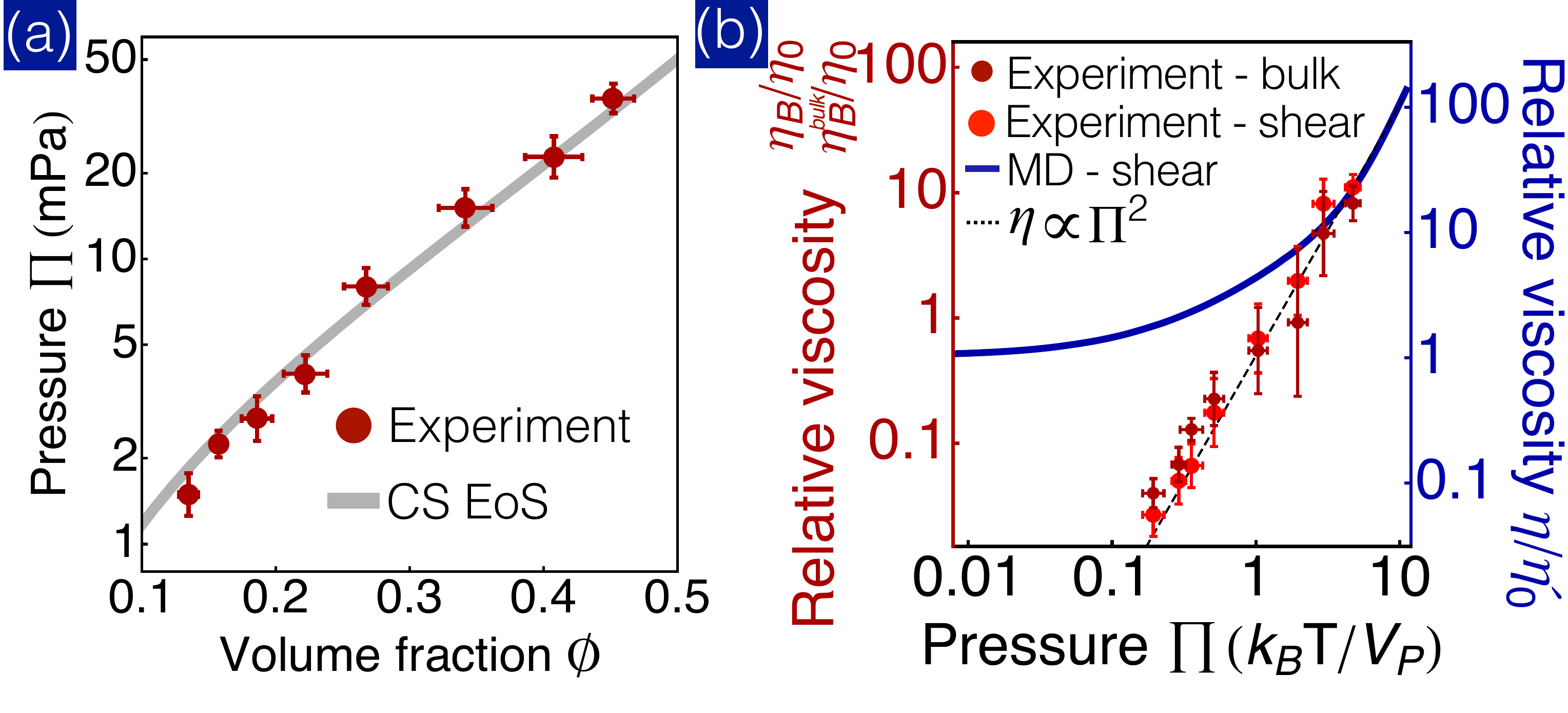}
\caption{(a) Osmotic pressure $\Pi$ (red disks) averaged over time is plotted versus $\phi$. We find that the measured trend $\Pi(\phi)$ is consistent with the Carnahan-Starling equation of state (CS EoS, gray line). (b) Log-log plot of the normalized Brownian contributions to the bulk $\eta_B^{bulk}/\eta_o$ (dark red points) and shear $\eta_B/\eta_o$ (light red points) viscosities versus $\Pi$ showing a $\Pi^2$ scaling (dashed black line), where $\eta_0$ is the solvent viscosity. The atomic liquid viscosity $\eta/\eta_0^\prime$ (blue curve, reproduced from~\cite{heyes2007system}) exhibits a similar scaling behavior only at high pressures, where $\eta_0^\prime$ is the ideal gas viscosity derived from the Boltzmann equation (see SI).}
\label{fig_4}
\end{figure}

In contrast to conventional mechanical measurements, which can only measure the flow-gradient stress and the difference between normal stresses, SALSA measures all stress components simultaneously. In Fig.~\ref{fig_4}(a) we report the pressure of the suspension at the eight volume fractions explored in Fig.~\ref{fig_3}. We find that the measured osmotic pressure arising from Brownian collisions (red disks) is well described by the Carnahan-Starling equation of state (gray line)~\footnote{We note that in previous confocal measurements where the pressure is determined by calculating the available volume to insert an additional sphere into a system, and its surface area~\cite{dullens2006direct}, the volume and corresponding surface area become exceedingly small and difficult to measure when suspensions are dense. This uncertainty results in a mismatch between data and theory. In our experiment, SALSA accurately captures the particle collision probabilities and correctly reports the pressure at all tested volume fractions.}. In addition, we find that both the shear ($\eta_B$, light red points) and bulk ($\eta_B^{bulk}$, dark red points) viscosities roughly exhibit $\Pi^2$ scaling (dashed black line), as shown in Fig.~\ref{fig_4}(b). While the underlying mechanism of such an empirical scaling remains an open question, we can qualitatively understand this scaling for the bulk viscosity using a dimensional analysis. Since the correlations in the Green-Kubo formula decay approximately exponentially in time, and the relaxation time $\tau$ does not increases significantly with increasing pressure over the range measured, we have
\begin{align*}
    \eta_B^{bulk} &\sim \int_0^{\infty} \langle (\Pi(t+\Delta t)-\bar\Pi) (\Pi(t) - \bar\Pi) \rangle d\Delta t \\
    &\sim \int_{0}^{\infty} C_{\Pi} e^{-\Delta t/\tau} d\Delta t  \sim \langle \Pi^2 \rangle - \langle \Pi \rangle^2
\end{align*}
where $C_{\Pi}$ is the variance of pressure~\cite{heyes1992molecular, fernandez2004molecular}.

While many previous studies have made analogies between the transport phenomena of colloidal systems and simple liquids~\cite{pham2002multiple, ediger1996supercooled, cheng2011imaging, lin2013far}, we find that the observed $\Pi^2$ scaling is actually absent in atomic systems. Specifically, the atomic viscosity (blue curve in Fig.~\ref{fig_4}(b)) exhibits a similar scaling behavior, but only at high pressures corresponding to large $\phi$. At low pressures, the viscosity trend deviates from the $\Pi^2$ scaling. We conjecture that this deviation is associated with the kinetic contribution to the viscosity ~\cite{chapman1970mathematical, hansen2013theory}, which is associated with atom velocity, insensitive to $\Pi$, and dominates in the dilute limit~(see SI). Collectively, our findings, which are made possible by SALSA, suggest that even the Brownian contribution to the colloidal viscosity can have a distinct transport mechanism than that in simple liquids.

In conclusion, we measure the stress fluctuation in colloidal liquids with SALSA, and experimentally demonstrate the well-known Green-Kubo relation~\cite{kubo1957statistical, green1954markoff}. Our measurements essentially show that ``as far as linear responses are concerned, the admittance is reduced to the calculation of time-fluctuations in equilibrium''~\cite{kubo1957statistical}. Previous pioneering experiments were able to combine the Green-Kubo relation with numerical simulations to extract the viscosity of a 2D dusty plasma~\cite{feng2011green}. These measurements, however, relied on assumptions for the interparticle potentials and ignored power-law decays in the stress correlation characteristic of 2D systems, which are known to lead to diverging integrals~\cite{alder1970decay, evans1980enhancedt, clercx1992brownian}. The analysis presented here avoids many of these complications and opens the door to further investigations of stress distributions in liquids under shear, confinement, and at high densities where the suspension becomes glassy~\cite{kegel2000direct, weeks2000three, hunter2012physics}. In such situations SALSA is still applicable since the solvent remains in equilibrium. More importantly, since the SALSA measurement is non-invasive, it also allows for probing the mechanical heterogeneity in a 3D colloidal glass~\cite{wyart2005geometric, schall2007structural, falk1998dynamics, bowick2016colloidal, chen2011measurement, tan2012understanding, chen2010low}, in which we can perform a time-average for particle-scale stress calculation. Measuring the temporal and spatial stress fluctuations in such a system will shed light on the generalization of the Green-Kubo relation in far-from-equilibrium systems and elucidate the mechanisms that underly the flow behaviors of disordered systems.

The authors thank James Sethna, Brian Leahy, James Swan, John Brady, and Wilson Poon for helpful discussions. I.C. and N.Y.C.L. gratefully acknowledge the Poon Laboratory at School of Physics \& Astronomy, University of Edinburgh for generous use of their PMMA suspensions. I.C. and N.Y.C.L. acknowledge funding from National Science Foundation (NSF) NSF CBET-PMP Award 1509308. M. B. was supported by Department of Energy DOE-DE-FG02-07ER46393 and continued support from NSF DMR-1507607.
%

\end{document}